\documentclass{article} \usepackage{spconf}

\usepackage{graphicx,psfrag}
\usepackage[cmex10]{amsmath}
\usepackage{amsfonts,amssymb,latexsym,bm}
\usepackage[hidelinks]{hyperref}
\usepackage{cite}
\usepackage{enumerate,subcaption}
\usepackage[usenames]{color}
\usepackage{stmaryrd}
\usepackage{footmisc}
\usepackage{xspace} 

\usepackage{algorithm,algpseudocode}

\algnewcommand{\Define}[1]{%
  \State \textbf{define:}
  \Statex \hspace*{\algorithmicindent}\parbox[t]{.8\linewidth}{\raggedright #1}
}
\algnewcommand{\Inputs}[1]{%
  \State \textbf{inputs:}
  \Statex \hspace*{\algorithmicindent}\parbox[t]{.8\linewidth}{\raggedright #1}
}
\algnewcommand{\Initialize}[1]{%
  \State \textbf{initialize:}
  \Statex \hspace*{\algorithmicindent}\parbox[t]{.8\linewidth}{\raggedright #1}
}

\usepackage{pbox}

\newcommand{\black}{\color{black}}

\renewcommand{\hat}{\widehat}
\renewcommand{\bar}{\overline}
\newcommand{\defn}{\triangleq}

 \newcommand{\mc}[1]{\ensuremath{\mathcal{#1}}}
\newcommand{\Real}{{\mathbb{R}}}
\newcommand{\Complex}{{\mathbb{C}}}

\newcommand{\tran}{^{\text{\textsf{T}}}}
\newcommand{\herm}{^{\text{\textsf{H}}}}

\DeclareMathOperator{\E}{\mathbb{E}}

\DeclareMathOperator{\tr}{tr}

\DeclareMathOperator{\prox}{prox}

\renewcommand{\eqref}[1]{(\ref{eq:#1})}

\newcommand{\Figref}[1]{Figure~\ref{fig:#1}}
\newcommand{\figref}[1]{Fig.~\ref{fig:#1}}

\renewcommand{\algref}[1]{Alg.~\ref{alg:#1}}
\newcommand{\lineref}[1]{line~\ref{line:#1}}

\newcommand{\Abf}{\bm{A}}

\newcommand{\gbf}{\bm{g}}

\newcommand{\rbf}{\bm{r}}
\newcommand{\ubf}{\bm{u}}
\newcommand{\vbf}{\bm{v}}

\newcommand{\xbf}{\bm{x}}

\newcommand{\ybf}{\bm{y}}

\newcommand{\rb}{\bm{r}}

\newcommand{\xb}{\bm{x}}

\newcommand{\iter}{t}
\newcommand{\iters}{{t+1}}

\newcommand{\itero}{{t-1}}

\newcommand{\MMSE}{_{\text{\sf MMSE}}}

\newcommand{\MAP}{_{\text{\sf MAP}}}

\renewcommand{\hat}[1]{\widehat{#1}}
\renewcommand{\bar}[1]{\overline{#1}}
\renewcommand{\vec}[1]{\boldsymbol{#1}}
\newcommand{\ovec}[1]{\bar{\boldsymbol{#1}}}
\newcommand{\hvec}[1]{\hat{\boldsymbol{#1}}}
\newcommand{\Jac}{\nabla}
\newcommand{\true}{_0}
\newcommand{\switch}{_\text{swi}}
\newcommand{\meas}{_\text{meas}}
\newcommand{\maxx}{_\text{max}}

\begin{document}
\setlength{\arraycolsep}{0.5mm}

\title{MRI Image Recovery using Damped Denoising Vector AMP}
\name{Subrata Sarkar$^\dagger$,
        Rizwan Ahmad$^*$, 
        and
        Philip Schniter$^\dagger$\thanks{This work was funded in part by the National Institutes of Health under grant R01HL135489 and by the National Science Foundation under grant CCF-1955587.}}
\address{$^\dagger$Dept. ECE, The Ohio State Univ., Columbus, OH, 43210, \{sarkar.51,\,schniter.1\}@osu.edu\\
         $^*$Dept. BME, The Ohio State Univ., Columbus, OH, 43210, rizwan.ahmad@osumc.edu}

\maketitle

\begin{abstract}
Motivated by image recovery in magnetic resonance imaging (MRI), 
we propose a new approach to solving linear inverse problems based on iteratively calling a deep neural-network, sometimes referred to as plug-and-play recovery.
Our approach is based on the vector approximate message passing (VAMP) algorithm, which is known for mean-squared error (MSE)-optimal recovery under certain conditions.
The forward operator in MRI, however, does not satisfy these conditions, and thus we design new damping and initialization schemes to help VAMP.
The resulting DD-VAMP++ algorithm is shown to outperform existing algorithms in convergence speed and accuracy when recovering images from the fastMRI database for the practical case of Cartesian sampling.
\black
\end{abstract}

\section{Introduction}
Magnetic resonance imaging (MRI) is a non-invasive diagnostic tool that provides excellent soft-tissue contrast without the use of ionizing radiation.
The primary drawback of MRI is that the data acquisition process is inherently slow. 
Because the scan time scales with the number of measurements, there is a motivation to collect as few measurements as possible while still providing accurate image recovery.
Although our work focuses on MRI, it has applicability to other inverse problems where the goal is to recover a richly structured signal from undersampled Fourier measurements.

\begin{figure}[t]
\centering
\newcommand{\sca}{0.22} 
\newcommand{\scb}{0.3} 
\newcommand{\sz}{0.8} 
    \begin{subfigure}[t]{\scb\columnwidth}
        \centering
        \psfrag{polynomial 25.07
        \includegraphics[scale=\sca,trim=0mm 0mm 0mm 0mm,clip]{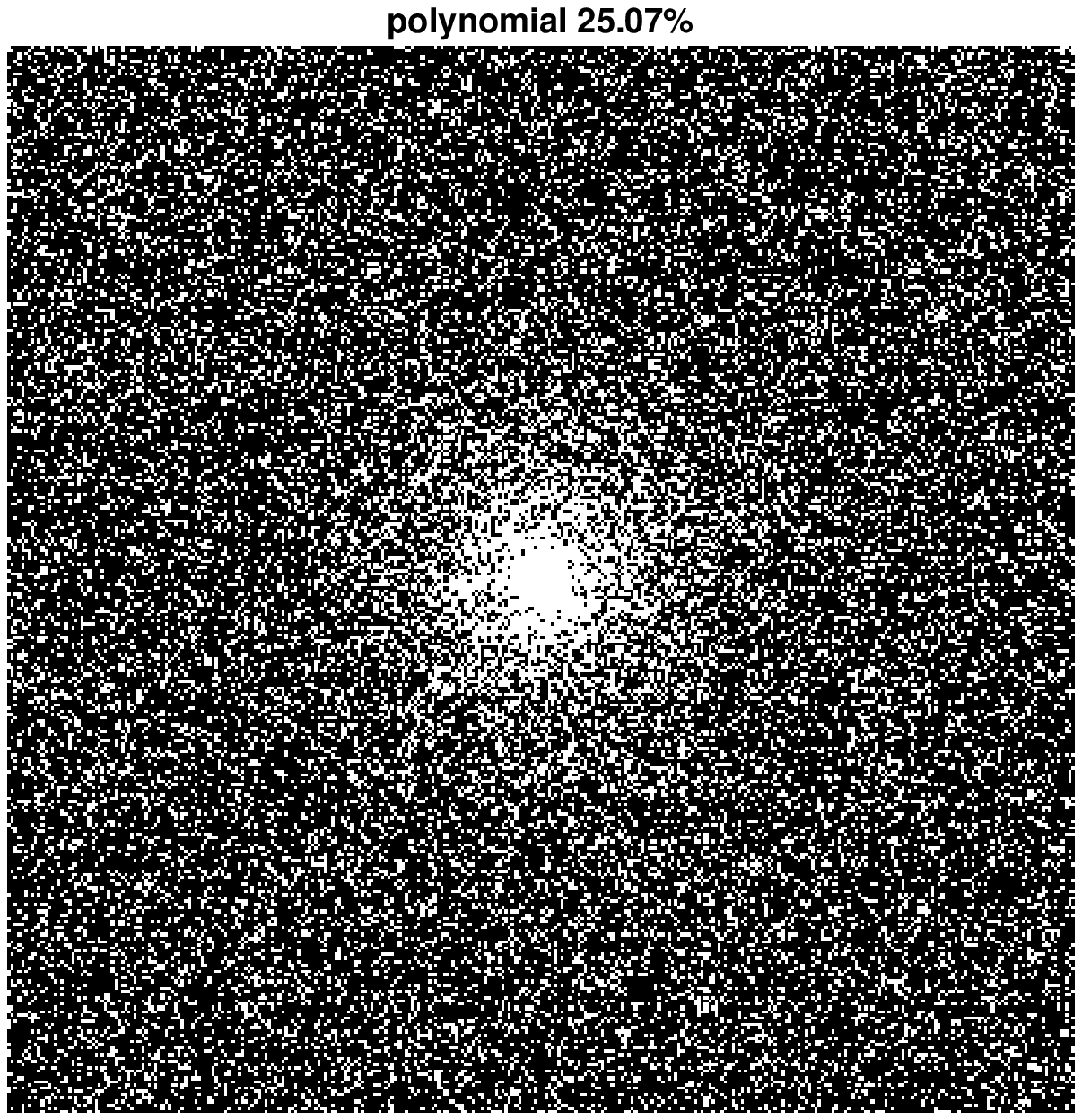}
    \end{subfigure}%
    \qquad
    \begin{subfigure}[t]{\scb\columnwidth}
        \centering
        \psfrag{cartesian 25
        \includegraphics[scale=\sca,trim=0mm 0mm 0mm 0mm,clip]{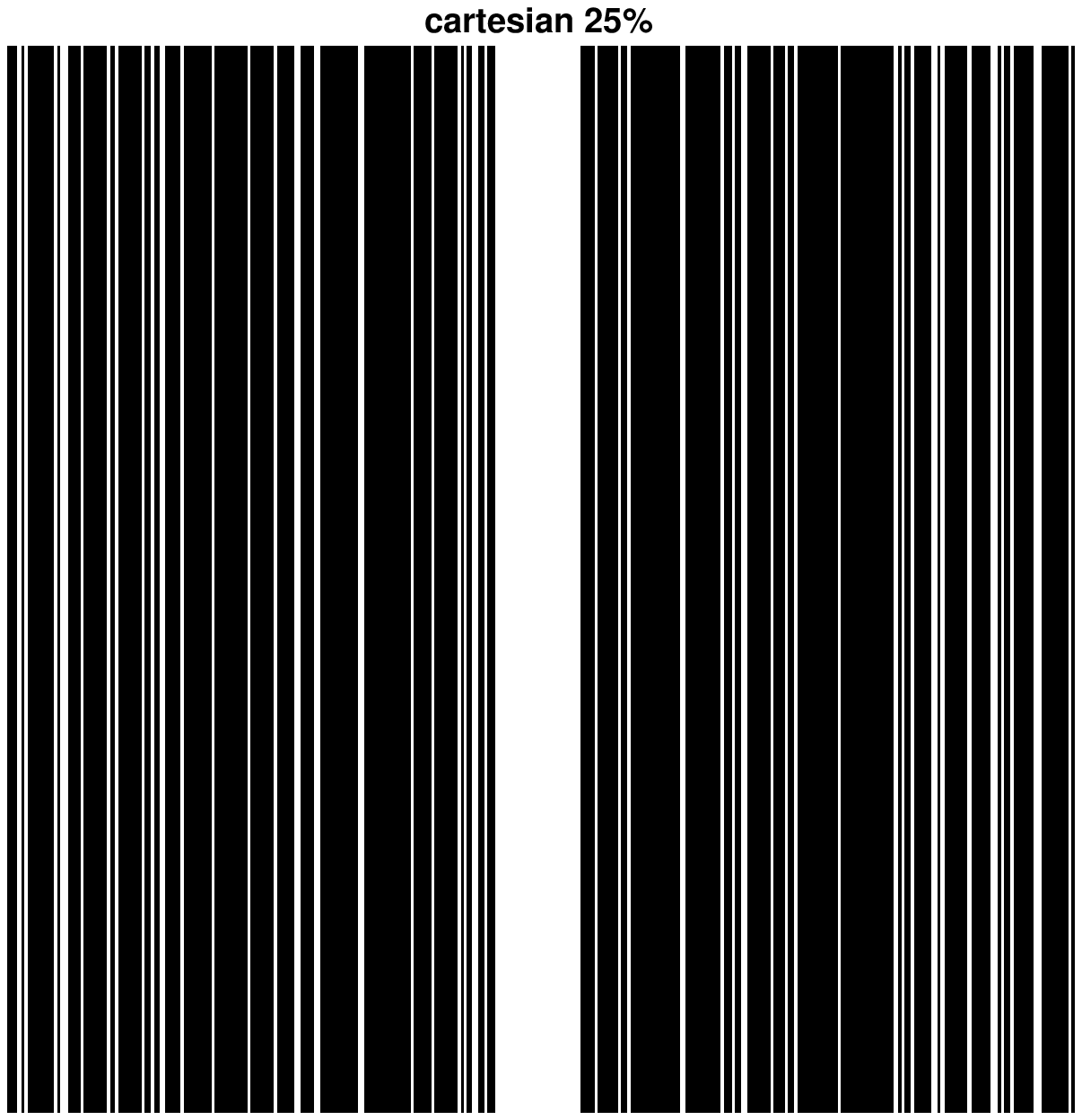}
    \end{subfigure}
    \caption{Typical sampling masks for acceleration $R=4$}
\label{fig:mri_masks}
\vspace{-5mm}
\end{figure}

In MRI, the Fourier-domain (or ``k space'') measurements $\vec{y}\in\Complex^M$ can be modeled as
\begin{align}
\vec{y} 
&= \vec{Ax}\true + \vec{w} \text{~~with~~}\vec{A}=\vec{MF}
\label{eq:y},
\end{align}
where 
$\vec{x}\true\in\Complex^N$ is the rasterized $N$-pixel image that we aim to recover,
$\vec{F}\in\Complex^{N\times N}$ is the unitary 2D discrete Fourier transform (DFT) matrix,
$\vec{M}\in\Complex^{M\times N}$ is a sampling mask (containing rows of the identity matrix $\vec{I}\in\Real^{N\times N}$), and 
$\vec{w} \sim \mc{N}(\vec{0}, \vec{I}/\gamma_w)$ is circularly symmetric additive white Gaussian noise (AWGN) of precision $\gamma_w$ (i.e., variance $1/\gamma_w$).
Typical sampling masks are illustrated in \figref{mri_masks}.
The ratio of pixels to measurements, $R\defn N/M$, is known as the ``acceleration rate.''
When $R>1$, there are many $\vec{x}$ that lead to the same $\vec{y}$, and so it is essential to use prior knowledge of $\vec{x}$ in recovery.

Many methods have been proposed for MRI image recovery.
As we describe in the next section, the classical approach is based on iterative optimization \cite{Lustig:SPM:08,Fessler:SPM:20}.
Recently, it has been proposed to train deep neural networks (DNNs) to directly map compressed measurements $\vec{y}$ to image estimates $\hvec{x}$, e.g., \cite{Jin:TIP:17,Zbontar:18}.
Although such DNNs are capable of excellent performance, training them requires huge fully-sampled k-space datasets (which may be difficult or impossible to obtain) and their performance can be degraded by changes in the sampling mask $\vec{M}$ between training and testing \cite{Ahmad:SPM:20}.

For these reasons, we focus on a hybrid approach where a DNN is iteratively called \emph{for image denoising}.
The DNN denoiser can be trained using relatively few images (via patches); no k-space data are needed.
Furthermore, the training of the denoiser is divorced from the sampling mask.
This hybrid approach is often referred to as ``plug-and-play'' recovery \cite{Venkatakrishnan:GSIP:13,Ahmad:SPM:20}.
Our approach is based on the vector AMP (VAMP) algorithm with two enhancements that lead to significant improvements in MRI image recovery: novel initialization and damping schemes.
Experiments with fastMRI \cite{Zbontar:18} knee images show advantages over existing plug-and-play approaches to MRI image recovery in speed and accuracy.

\section{Background}

\subsection{Optimization-based methods}

The conventional approach to MRI image recovery \cite{Lustig:SPM:08,Fessler:SPM:20} is to pose and solve an optimization problem of the form
\begin{align}
\hvec{x} 
&= \arg\min_{\vec{x}} 
\big\{ \tfrac{\gamma_w}{2}\|\vec{y} - \vec{Ax}\|^2 + \phi(\vec{x}) \big\}
\label{eq:mri_optim},
\end{align}
where the regularizer $\phi:\Complex^N\rightarrow\Real_+$ aims to penalize $\vec{x}$ that are not valid images.
It is common to choose $\phi(\vec{x})=\lambda\|\vec{\Psi x}\|_1$ with a suitable (e.g., wavelet) transform $\vec{\Psi}$ and carefully chosen $\lambda>0$.
In this case, the regularization encourages sparsity in the wavelet coefficients $\vec{\Psi x}$.

Many algorithms have been designed to solve optimization problems of the form \eqref{mri_optim} with convex $\phi(\cdot)$ \cite{Fessler:SPM:20}.
A famous one is the alternating directions method of multipliers (ADMM) \cite{Boyd:FTML:11}, which for \eqref{mri_optim} iterates
\begin{subequations}
\label{eq:admm}
\begin{eqnarray}
\xbf^\iters &=& \arg\min_{\xbf} \big\{ \tfrac{\gamma_w}{2} \|\ybf - \Abf\xbf\|^2 + \tfrac{\gamma}{2}\|\xbf-\vbf^\iter + \ubf^\iter\|^2 \big\}
\qquad
\label{eq:admm_loss}\\
\vbf^\iters &=& \prox_{\gamma^{-1}\phi}(\xbf^\iters+\ubf^\iter)
\label{eq:admm_prox}\\
\ubf^\iters &=& \ubf^\iter + \left(\xbf^\iters - \vbf^\iters\right) 
\label{eq:admm_dual},
\end{eqnarray}
\end{subequations}
for $t=0,1,2,\dots$, starting from $\vec{v}^0=\vec{0}=\vec{u}^0$, where 
\begin{align}
\prox_{\rho}(\rbf) 
\defn \arg\min_{\xbf} \big\{ \rho(\xbf) + \tfrac{1}{2}\|\xbf-\rbf\|^2 \big\}
\label{eq:prox}.
\end{align}
In \eqref{admm}, $\gamma>0$ is a tunable stepsize that affects the speed of ADMM's convergence but not its fixed point.

\subsection{Plug-and-play methods}

It can be useful to interpret the optimization \eqref{mri_optim} as maximum a posteriori (MAP) estimation of $\vec{x}$. 
This can be seen using Bayes rule, $p(\xbf|\ybf)=p(\ybf|\xbf)p(\vec{x})/p(\vec{y})$, as follows,
\begin{align}
\hat{\xbf}\MAP
&\defn \arg\max_{\xbf} p(\xbf|\ybf) \\
&= \arg\min_{\xbf} \big\{ -\ln p(\ybf|\xbf) - \ln p(\xbf) \big\}
\label{eq:mri_map},
\end{align}
and confirming that \eqref{mri_map} matches \eqref{mri_optim} under 
the prior 
$p(\vec{x})\propto e^{-\phi(\vec{x})}$ 
and likelihood 
$p(\vec{y}|\vec{x})=\mc{N}(\vec{y};\vec{Ax},\vec{I}/\gamma_w)$.
Similarly, ADMM's prox step \eqref{admm_prox} can be interpreted as MAP denoising, i.e.,
MAP estimation of $\vec{x}\sim p(\vec{x})\propto e^{-\phi(\vec{x})}$ from the noise-corrupted measurement $\vec{r}\!=\!\vec{x}\!+\!\vec{w}$ with $\vec{w}\!\sim\!\mc{N}(\vec{0},\vec{I}/\gamma)$.

Leveraging this MAP denoising interpretation, Bouman et al.\ \cite{Venkatakrishnan:GSIP:13} proposed to replace $\prox_{\gamma^{-1}\phi}$ in line \eqref{admm_prox} with a call to a black-box image denoiser $\vec{f}:\Complex^N\rightarrow\Complex^N$ like BM3D~\cite{Dabov:TIP:07} or DnCNN~\cite{Zhang:TIP:17}, and coined the approach ``plug-and-play'' (PnP) ADMM. 
In MRI, PnP methods tend to significantly outperform \cite{Ahmad:SPM:20} the conventional optimization approach \eqref{mri_optim} to image recovery.
Note, however, that when \eqref{admm_prox} is replaced with a denoising step of the form $\vec{v}^\iters=\vec{f}(\vec{x}^\iters\!+\!\vec{u}^\iter)$, the stepsize $\gamma$ \emph{does} affect the fixed-point and thus must be tuned.


\subsection{Approximate message passing}

Approximate message passing (AMP) \cite{Donoho:PNAS:09} is another iterative approach to computing the solution of \eqref{mri_optim} or \eqref{mri_map}, i.e., the MAP estimate of $\vec{x}$ from $\vec{y}$.
But it can also \cite{Donoho:ITW:10a} be used to approximate $\hvec{x}\MMSE\defn\E\{\vec{x}\,|\,\vec{y}\}$, the minimum mean-squared error (MMSE) estimate of $\vec{x}$ from $\vec{y}$.  
In general, AMP iterates 
\begin{subequations}
\label{eq:amp}
\begin{eqnarray}
\vec{v}^\iters 
&=& \beta\cdot\big( \vec{y}-\vec{Ax}^\iter 
        + \tfrac{1}{M}\vec{v}^\iter \tr\{\Jac\vec{f}^\iter(\vec{x}^\itero\!+\!\vec{A}\herm\vec{v}^\iter)\} \big)
\qquad
\label{eq:amp_onsager}\\
\tau^\iters
&=& \tfrac{1}{M}\|\vec{v}^\iters\|^2
\label{eq:amp_noisevar}\\
\vec{x}^\iters
&=&\vec{f}^\iters(\vec{x}^\iter + \vec{A}\herm\vec{v}^\iters)
\end{eqnarray}
\end{subequations}
over $t=0,1,2,\dots$, starting from $\vec{v}^0=\vec{0}=\vec{x}^0$, where 
$\vec{f}^\iter(\cdot)$ is a Lipschitz denoising function,
$\tr\{\Jac\vec{f}^\iter(\vec{r})\}$ is the trace of the Jacobian of $\vec{f}^\iter$ at $\vec{r}$,
and $\beta=N/\|\vec{A}\|_F^2$.
When configured for MAP estimation, AMP uses the MAP denoiser $\vec{f}\MAP^\iter(\vec{r})=\prox_{\tau^\iter\phi}(\vec{r})$.
When configured for MMSE estimation, AMP instead uses the MMSE denoiser $\vec{f}\MMSE^\iter(\vec{r})=\E\{\vec{x}\,|\,\vec{r}\}$ for $\vec{r}=\vec{x}+\vec{w}$ with $\vec{w}\sim\mc{N}(\vec{0},\tau^t\vec{I})$.

When the forward operator $\vec{A}$ is large and i.i.d.\ sub-Gaussian, the macroscopic behavior of AMP is rigorously characterized by a scalar state-evolution 
for any Lipschitz $\vec{f}^\iter$ \cite{Bayati:TIT:11,Berthier:II:19}.
When $\vec{f}^t$ is the MMSE denoiser and the state-evolution has a unique fixed point, AMP provably converges to the MMSE-optimal estimate $\hvec{x}\MMSE$ \cite{Bayati:TIT:11,Berthier:II:19}.
Although an exact MMSE denoiser for images is unknown, one could instead use a black-box approximation (like BM3D) within AMP, as proposed in \cite{Metzler:ICIP:15} and called ``denoising-AMP'' (D-AMP).
In that setting, the Jacobian term in \eqref{amp_onsager} is typically approximated using the approach from \cite{Ramani:TIP:08},
\begin{align}
\tr\{\Jac\vec{f}^\iter(\vec{r})\}
\approx \epsilon^{-1}\vec{q}\herm\big[\vec{f}^\iter(\vec{r}+\epsilon\vec{q})-\vec{f}^\iter(\vec{r)}\big]
\label{eq:trJfapprox} ,
\end{align}
for random $\vec{q}\sim\mc{N}(\vec{0},\vec{I})$ and small positive $\epsilon$.

For AMP to behave as expected, the forward operator $\vec{A}$ should resemble a typical realization of a large, i.i.d.\ sub-Gaussian matrix.
To expand the applicability of AMP, a variation called ``vector AMP (VAMP)'' was proposed \cite{Rangan:TIT:19}.
VAMP has essentially the same desirable properties of AMP (e.g., rigorous state evolution and provable MMSE denoising) but they hold for a much larger class of random matrices: right orthogonally invariant (ROI) ones \cite{Rangan:TIT:19,Fletcher:NIPS:18}.
An ROI matrix has a singular value decomposition of the form $\vec{USV}\herm$, where $\vec{U}$ is unitary, $\vec{S}$ is diagonal, and $\vec{V}$ is Haar (i.e., uniformly distributed over the set of unitary matrices). 
A denoising VAMP (D-VAMP) was proposed in \cite{Schniter:BASP:17} that uses a black-box image denoiser $\vec{f}^t$ together with the Jacobian approximation \eqref{trJfapprox}.

\subsection{AMP for MRI}

While the PnP-ADMM algorithm requires manual tuning of both the ADMM parameter $\gamma$ and the noise variance assumed by the denoiser $\vec{f}$,
the AMP algorithm \eqref{amp} is self-tuning; it estimates the denoiser-input variance $\tau^\iter$ at each iteration $t$. 
This motivates the application of AMP to MRI image recovery. 

Unfortunately, neither AMP nor VAMP 
works as expected in MRI because $\vec{A}$ in \eqref{y} is neither i.i.d.\ nor ROI.
In fact, MRI applications of AMP or VAMP often diverge.
For this reason, several MRI-specific variations of AMP have been proposed.
For example, \cite{Eksioglu:JIS:18} used D-AMP with $\beta=1$ instead of the default value $\beta=N/\|\vec{A}\|_F^2=N/M$, which acts to slow down the algorithm to help convergence but degrades the fixed points. 
A refinement based on a matrix-valued $\beta$ was proposed in \cite{Sarkar:Diss:20}.
In \cite{Millard:20}, a hybrid between AMP and VAMP was proposed for the special case of point sampling (recall \figref{mri_masks}), called variable density (VD)-AMP.
We numerically investigate these approaches in the sequel.

\section{Proposed Approach}

\subsection{Damped Denoising VAMP}

The proposed damped D-VAMP (DD-VAMP) algorithm is summarized in \algref{DDVAMP}. 
Note that, in the absence of damping (i.e., $\theta\!=\!1\!=\!\zeta^t$), DD-VAMP reduces to D-VAMP from \cite{Schniter:BASP:17}.
An explanation of each step is now provided.

Line~\ref{line:x2_iter_VP} of \algref{DDVAMP} computes the MMSE/MAP estimate of $\vec{x}$ under 
the likelihood function $\mc{N}(\vec{y};\vec{Ax},\vec{I}/\gamma_w)$
and the pseudo-prior $\vec{x}\sim\mc{N}(\vec{x};\vec{r}_2^\iter,\vec{I}/\gamma_2^\iter)$
via 
\begin{align}
\lefteqn{
\vec{g}(\vec{r};\gamma)
\defn \arg\min_{\xbf} \big\{ \tfrac{\gamma_w}{2} \|\ybf - \Abf\xbf\|^2 + \tfrac{\gamma}{2}\|\xbf-\rbf\|^2 \big\} }
\label{eq:g}\\
&= \vec{F}\herm(\gamma_w\vec{M}\tran\vec{M}+\gamma\vec{I})^{-1}
       (\gamma \vec{Fr} + \gamma_w\vec{M}\tran\vec{y}) 
\label{eq:g_singlecoil} ,
\end{align}
where \eqref{g} holds for general $\vec{A}$ and \eqref{g_singlecoil} holds under \eqref{y}.
Because $\vec{M}$ is a masking matrix, the computational complexity of \eqref{g_singlecoil} is dominated by two FFTs.
Line~\ref{line:alf2_iter_VP} computes the sensitivity of this linear estimator $\vec{g}$ as 
\begin{align}
\tr\{\Jac\vec{g}(\vec{r};\gamma)\}/N
&= \gamma \tr\{(\gamma_w\vec{A}\herm\vec{A}+\gamma\vec{I})^{-1}\}/N
\label{eq:trJg}\\
&= \big((1-\tfrac{M}{N})\gamma_w+\gamma\big)/\big(\gamma_w+\gamma\big)
\label{eq:trJg_singlecoil} ,
\end{align}
where \eqref{trJg} holds for general $\vec{A}$ and \eqref{trJg_singlecoil} holds under \eqref{y}.
Lines~\ref{line:r1_iter_VP}--\ref{line:gam1_iter_VP} perform ``Onsager cancellation,'' which produces the $(\vec{r}_1^\iter,\gamma_1^\iter)$ passed on to the denoising stage. 
If $\vec{f}^\iter$ is Lipschitz and $\vec{A}$ is large and ROI, then these latter terms obey the statistical model 
$\vec{r}_1^\iter=\vec{x}\true+\mc{N}(\vec{0},\vec{I}/\gamma_1^\iter)$ 
\cite{Fletcher:NIPS:18}.

Lines~\ref{line:x1_iter_VP}--\ref{line:gam2_iter_VP} of \algref{DDVAMP} comprise the denoising stage, which calls the denoiser $\vec{f}^\iter$ and approximates its sensitivity $\tr\{\Jac\vec{f}^\iter(\vec{r}_1^\iter)\}/N$ in a manner similar to D-AMP (see \eqref{trJfapprox}).
Note that the denoiser can leverage knowledge of the noise variance $\tau^t=1/\gamma_1^t$.
Lines~\ref{line:r2_iter_VP}--\ref{line:gam2_iter_VP} perform Onsager cancellation, producing $(\ovec{r}_2^\iters,\bar{\gamma}_2^\iters)$ for use by the linear stage.

\begin{algorithm}[t]
\caption{Damped Denoising-VAMP (DD-VAMP)}
\label{alg:DDVAMP}
\begin{algorithmic}[1]
\Require $\rb_{2}^0\in\Complex^N,~~
\gamma_{2}^0>0,~~
\theta\in(0,1],~~
\vec{q}\sim\mc{N}(\vec{0},\vec{I})$
\For{$t=0,\dots,T_{\max}$}
\\ ~~Linear Stage:
\State ${\xb}_{2}^{\iter}=\gbf(\rb_{2}^{\iter};{\gamma}_{2}^{\iter})$\label{line:x2_iter_VP}
\State $\alpha_2^{\iter}= \tr\{\Jac \gbf(\rb_{2}^{\iter};\gamma_{2}^{\iter})\}/N$\label{line:alf2_iter_VP}
\State ${\rb}_{1}^{\iter}=(\xb_{2}^{\iter}-\alpha_2^\iter\rb_{2}^\iter) /(1-\alpha_{2}^{\iter})\label{line:r1_iter_VP}$
\State ${\gamma}_{1}^{\iter}= \gamma_2^\iter (1-\alpha_2^\iter)/\alpha_2^\iter\label{line:gam1_iter_VP}$ 
\\~~Denoising:
\State ${\xb}_{1}^{\iter}=\vec{f}^\iter(\rb_{1}^\iter)$\label{line:x1_iter_VP}
\State $\bar{\alpha}_1^\iter = \epsilon^{-1}\vec{q}\herm\big[\vec{f}^\iter(\vec{r}_1^\iter+\epsilon\vec{q})-\vec{f}^\iter(\vec{r}_1^\iter)\big]$  \label{line:alf1_iter_VP}
\State $\alpha_{1}^{\iter}= \{\theta(\bar{\alpha}_1^\iter)^{\frac{1}{2}} + (1-\theta)(\alpha_1^\itero)^{\frac{1}{2}}\}^2$ ~~(damping)\label{line:alf1_damp_VP}
\State $\bar{\rb}_{2}^{\iters}=(\xbf_1^\iter - \alpha_1^\iter\rbf_1^\iter)/(1-\alpha_1^\iter)$  \label{line:r2_iter_VP} 
\State $\bar{\gamma}_2^\iters = \gamma_1^\iter (1 - \alpha_1^\iter) / \alpha_1^\iter$  \label{line:gam2_iter_VP}
\\~~Damping:
\State Choose damping factor $\zeta^t\in(0,1]$\label{line:zeta_VP}
\State $\rbf_2^\iters = \zeta^\iter\bar{\rbf}_2^\iters + (1-\zeta^\iter)\rbf_2^\iter$\label{line:r2_damp_VP} 
\State $\gamma_2^\iters = \{\zeta^\iter(\bar{\gamma}_2^\iters)^{-\frac{1}{2}} + (1-\zeta^\iter)(\gamma_2^\iter)^{-\frac{1}{2}}\}^{-2}$\label{line:gam2_damp_VP}
\EndFor 
\end{algorithmic}
\end{algorithm}

Before $(\ovec{r}_2^\iters,\bar{\gamma}_2^\iters)$ are passed on, however, they are damped in lines~\ref{line:r2_damp_VP}--\ref{line:gam2_damp_VP} using the damping factor $\zeta^\iter\in(0,1]$ chosen in \lineref{zeta_VP};
with $\zeta^\iter=1$ there would be no damping, and as $\zeta^\iter$ decreases there would be more damping.
A similar form of damping is performed on $\alpha_1^\iter$ in \lineref{alf1_damp_VP} using the factor $\theta$.
Intuitively, the goal of damping is to slow down VAMP without changing its fixed points.
The $\vec{r}^\iter_2$-damping in \lineref{r2_damp_VP} was shown to be sufficient to ensure the convergence of a double-loop version of VAMP in the strongly convex scenario \cite{Fletcher:GEC}.
But additional damping of $\gamma_2^\iter$ and $\alpha_1^\iter$ is needed to stabilize D-VAMP for MRI image recovery.

Although $\gamma_i$-damping was suggested in \cite[eq.(27)]{Rangan:TIT:19}, the approach here is different in that damped quantities are first converted to amplitudes.
That is, for a variance term like $\alpha_1$ we damp the square-root, and for a precision term like $\gamma_2$ we damp the inverse square root.
The $\alpha_1$-damping is new and motivated by the fact that the Jacobian approximation in \lineref{alf1_iter_VP} can be very ``noisy,'' especially in the first iterations. 
Although it could be improved using more averaging, i.e., 
\begin{align}
\tr\{\Jac\vec{f}(\vec{r})\}
\textstyle 
\approx \frac{1}{K}\sum_{k=1}^K \epsilon^{-1}\vec{q}_k\herm\big[\vec{f}(\vec{r}+\epsilon\vec{q}_k)-\vec{f}(\vec{r)}\big]
\label{eq:trJfapprox2} ,
\end{align}
with i.i.d.\ $\{\vec{q}_k\}_{k=1}^K$ and large $K$, this would require $K\!+\!1$ denoiser-calls per VAMP iteration and thus be too expensive.

There are several ways to choose the damping factor $\zeta^\iter$ in \lineref{zeta_VP}.
One is to use the rule
\begin{eqnarray}
\zeta^\iter 
&=& \frac{2}{1+\max\{\gamma_1^\iter/\bar{\gamma}_2^\iters, 
                    \bar{\gamma}_2^\iters/\gamma_1^\iter\}}
= 2\min\{\alpha_1^\iter,\alpha_2^\iter\}
\label{eq:vamp_zeta},
\quad
\end{eqnarray}
which is sufficient in the double-loop strongly convex scenario \cite{Fletcher:GEC}.
Another is to chose some fixed $\zeta^\iter\!=\!\zeta$ for all $t$.

\subsection{ADMM-PR and DD-VAMP++}

Empirically, we find that DD-VAMP converges to fixed points that are similar or better than those of PnP-ADMM. 
However, we find that it can converge somewhat slowly as a result of damping.
To circumvent this issue, we propose a carefully initialized version called ``DD-VAMP++.'' 
The challenge with initializing DD-VAMP, i.e., choosing $(\vec{r}_2^0,\gamma_2^0)$, is that the two values must be consistent.
That is, if $\vec{r}_2^0$ is very good (i.e., $\approx \vec{x}\true$) but $\gamma_2^0$ is not matched to $\vec{r}_2^0$, then DD-VAMP will initially move $\vec{r}_2^\iter$ away from $\vec{r}_2^0$ while adjusting $\gamma_2^\iter$, and finally return $\vec{r}_2^\iter$ to the neighborhood of $\vec{r}_2^0$.

The proposed initialization approach exploits the fact that VAMP is closely related to the Peaceman-Rachford (PR) variant \cite{Gabay:Chap:83} of ADMM, which is the iteration \eqref{admm} with an additional $\vec{u}$-update of the form \eqref{admm_dual} between \eqref{admm_loss} and \eqref{admm_prox}.
In particular, when VAMP is run with fixed $\gamma_1\!=\!\gamma_2\!=\!\gamma$ for all iterations $t$, it reduces to ADMM-PR
\cite{Fletcher:GEC}.

For DD-VAMP++, we first run PnP-ADMM-PR for $T\switch$ iterations, 
starting from $\vec{u}^0=\vec{0}=\vec{v}^0$ and a well-chosen value of $\gamma$,
and then switch to DD-VAMP, simply by allowing the values of $\gamma_1$ and $\gamma_2$ to update according to \algref{DDVAMP}.
DD-VAMP++'s two adjustable parameters, $T\switch$ and $\gamma$, can be tuned using training data, as described in the next section.

\section{Numerical Experiments}\label{sec:mri_sim}

We now present numerical experiments that compare the proposed DD-VAMP and DD-VAMP++ algorithms to the existing PnP-ADMM \cite{Venkatakrishnan:GSIP:13}, VD-AMP \cite{Millard:20}, and BM3D-AMP-MRI \cite{Eksioglu:JIS:18} algorithms for MRI image recovery. 
For completeness, we also test the PnP-ADMM-PR algorithm, which is run during the initialization phase of DD-VAMP++.

For the DD-VAMP, DD-VAMP++, PnP-ADMM, and PnP-ADMM-PR algorithms, we used the DNN-based denoiser DnCNN \cite{Zhang:TIP:17}. 
(The BM3D denoiser \cite{Dabov:TIP:07} yields similar image quality but runs significantly slower.)
Our D-VAMP algorithms used the DnCNN implementation in the D-AMP toolbox
{\small \href{https://github.com/ricedsp/D-AMP_Toolbox}{\tt https://github.com/ricedsp/D-AMP\_Toolbox}},
\linebreak
which calls an instance of DnCNN trained with a noise variance close to the requested one, $\tau^\iter=1/\gamma_1^\iter$.
VD-AMP and BM3D-AMP-MRI were both implemented using the respective groups' code.
To denoise, VD-AMP uses SURE-tuned wavelet thresholding while BM3D-AMP-MRI uses a modified BM3D.
All experiments used mid-slice, non-fat-suppressed, 128$\times$128 knee images from fastMRI \cite{Zbontar:18}.
We focused on the Cartesian mask (see \figref{mri_masks}) at acceleration $R\!=\!N/M\!=\!4$.
The variance of the AWGN $\vec{w}$ in \eqref{y} was adjusted to achieve an SNR, $\|\vec{Ax}\|^2/\|\vec{w}\|^2$, of 40 dB.

The PnP-ADMM, PnP-ADMM-PR, DD-VAMP, and DD-VAMP++ algorithms (all initialized at zero) have tuning parameters that must be optimized for best performance.
For PnP-ADMM and PnP-ADMM-PR we tuned $\gamma$ in \eqref{admm},
for DD-VAMP we tuned $\gamma_2^0$, and
for DD-VAMP++ we tuned $\gamma$ and the switching time $T\switch$.
In all cases, these parameters were tuned over a grid using the following procedure.
For each hypothesized value, the algorithm was used to recover 30 training images, the recovery-NMSE was averaged over iterations $T\meas=$~35 to $T\maxx=$~150, and the median was taken over the training images (to reduce the effect of outlier images).
Finally, the parameter value that minimized the median NMSE was selected.

\begin{figure}[t]
    \psfrag{iterations1}[b][b][0.7]{\sf iteration}
    \psfrag{iterations2}[b][b][0.7]{\sf iteration}
    \psfrag{median NMSE on training images}[b][b][0.7]{\sf median NMSE on training set}
    \psfrag{median NMSE on test images}[b][b][0.7]{\sf median NMSE on test set}
    \psfrag{nmse}[b][b][0.7]{\sf NMSE}
    \includegraphics[width=\columnwidth]{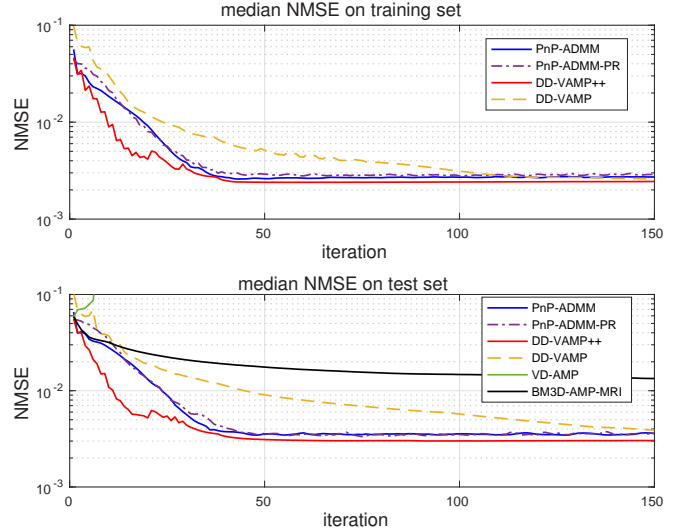}
    \caption{Image recovery NMSE versus iteration.}
    \label{fig:traj_cart}
\end{figure}

\Figref{traj_cart} shows NMSE $\defn\|\hvec{x}-\vec{x}\true\|^2/\|\vec{x}\true\|^2$ versus iteration with a Cartesian mask, which is the mask most commonly used in clinical practice.
The top subplot shows median NMSE over the 30 training images for the tuned version of each algorithm.
There, DD-VAMP++ converged fastest and to the best NMSE.
DD-VAMP achieved the 2nd-best NMSE at 150 iterations, but converged slowly.
The two versions of PnP-ADMM showed similar behavior, although PnP-ADMM-PR was less smooth and its final NMSE was slightly worse.
The bottom subplot of \figref{traj_cart} shows median NMSE over the 19 remaining fastMRI mid-slice non-fat-suppressed images, which were used as a test set.
Here too, we see that DD-VAMP++ converged fastest to the best NMSE, with PnP-ADMM and PnP-ADMM-PR close behind.
After 150 iterations, DD-VAMP is still converging to what seems to be an excellent terminal NMSE.
BM3D-AMP-MRI converged to a significantly worse NMSE; we conjecture that this was due to its heuristic $\beta$ modification and the mismatch between Cartesian-masked $\vec{A}$ and typical i.i.d.\ sub-Gaussian matrices.
Finally, VD-AMP (which was designed specifically for point masks) diverged; it is apparently not compatible with Cartesian masks. 
In experiments that we conducted with point masks (not shown here for reasons of space), VD-AMP converged very quickly to an NMSE $>$10~dB worse than DD-VAMP++ and PnP-ADMM, but
consistent with the results in \cite{Millard:20}.
Evidently, VD-AMP's wavelet-thresholding denoiser is no match for BM3D or DnCNN.

%

\begin{figure}[t]
\newcommand{\sca}{0.22}
\newcommand{\scb}{0.3}
\newcommand{\sz}{0.55}
\centering
    \begin{subfigure}[t]{\scb\columnwidth}
        \centering
        \psfrag{original}[b][b][\sz]{\sf original}
        \includegraphics[scale=\sca,trim=0mm 0mm 0mm 0mm,clip]{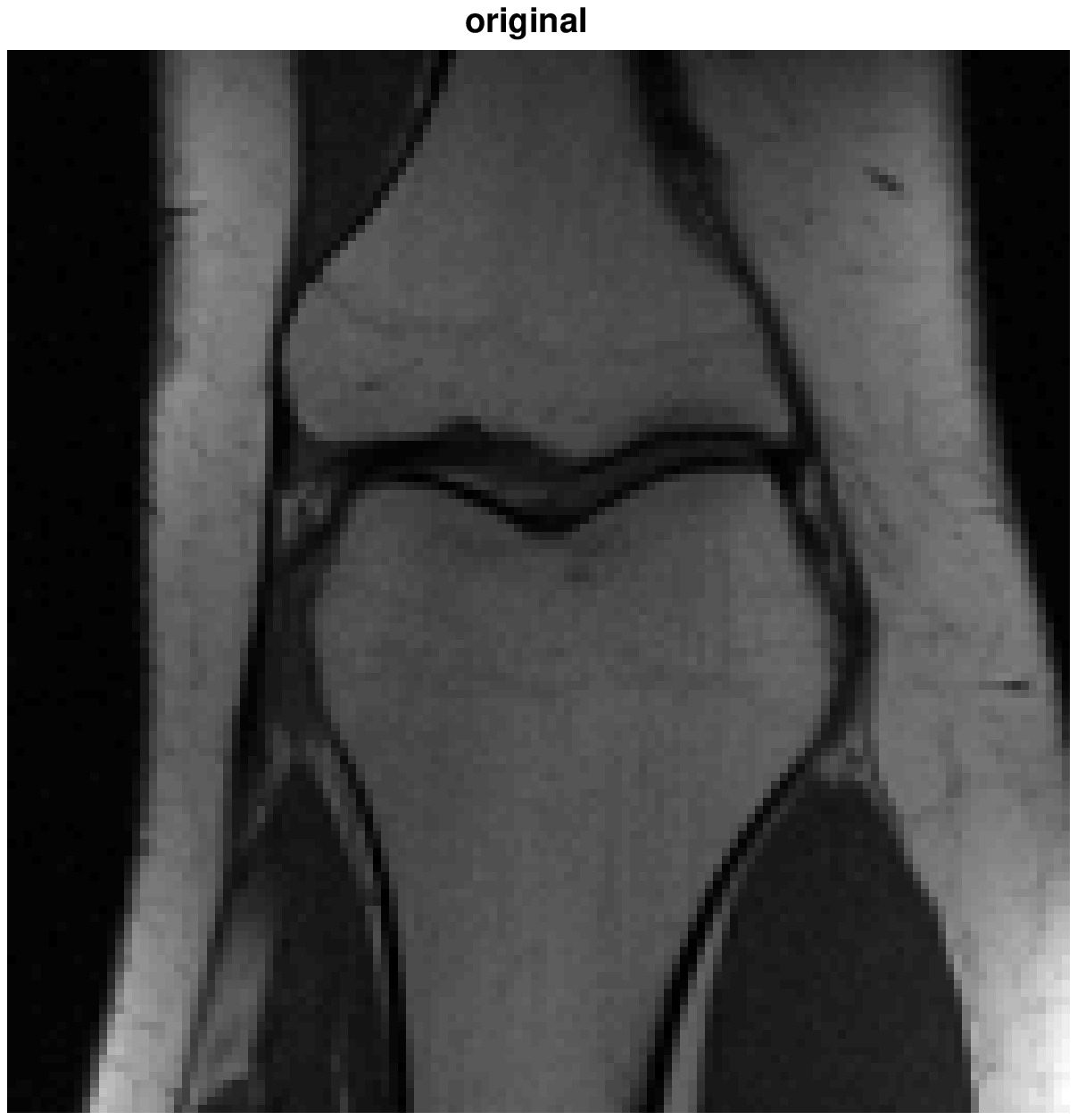}
    \end{subfigure}
    ~
    \begin{subfigure}[t]{\scb\columnwidth}
        \centering
        \psfrag{ADMM: -24.443, 0.928}[b][b][\sz]{\sf ADMM: -24.443, 0.928}
        \includegraphics[scale=\sca,trim=0mm 0mm 0mm 0mm,clip]{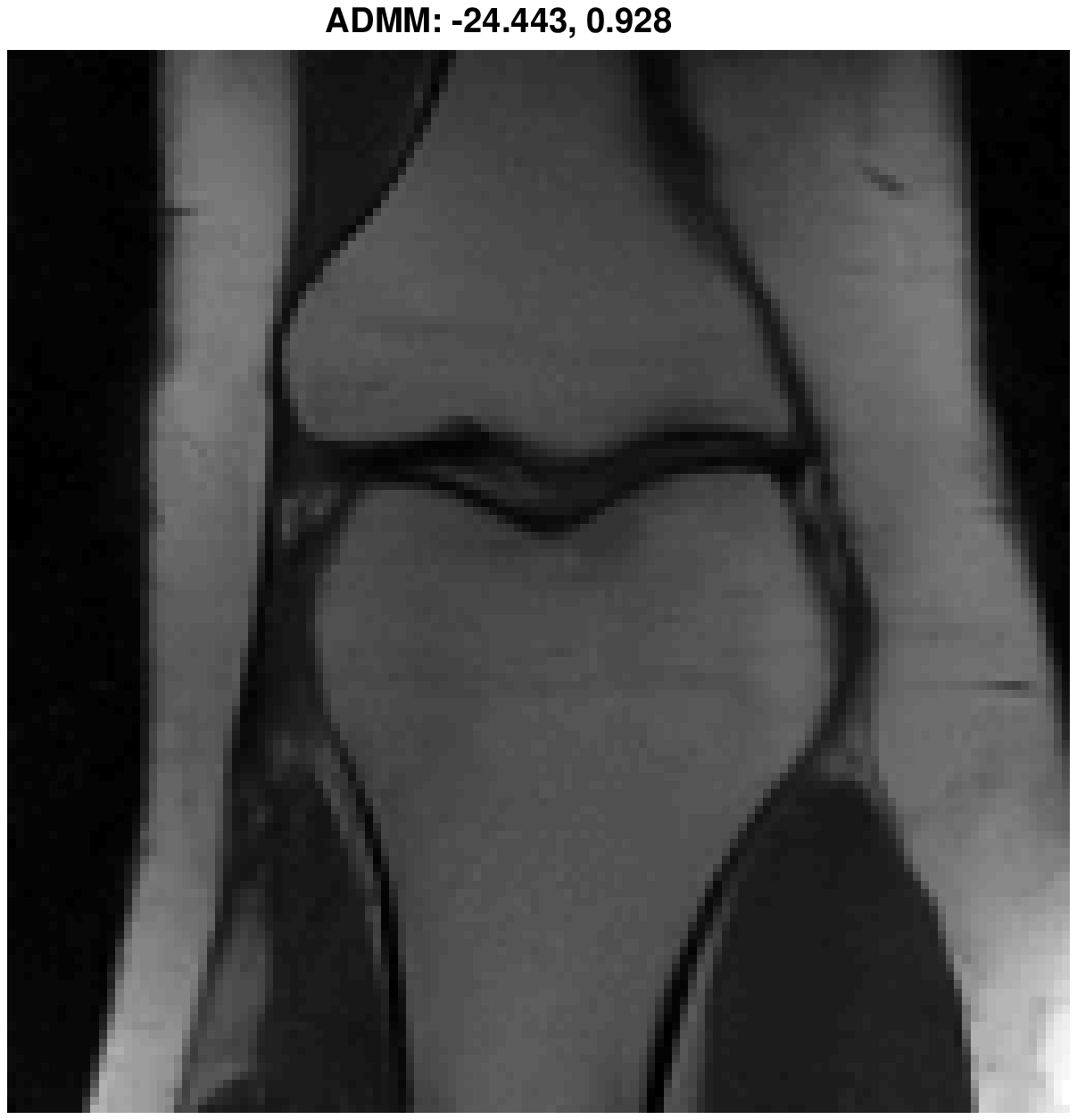}
    \end{subfigure}
    ~
    \begin{subfigure}[t]{\scb\columnwidth}
        \centering
        \psfrag{ADMM-PR: -24.694, 0.927}[b][b][\sz]{\sf \quad ADMM-PR: -24.694, 0.927}
        \includegraphics[scale=\sca,trim=0mm 0mm 0mm 0mm,clip]{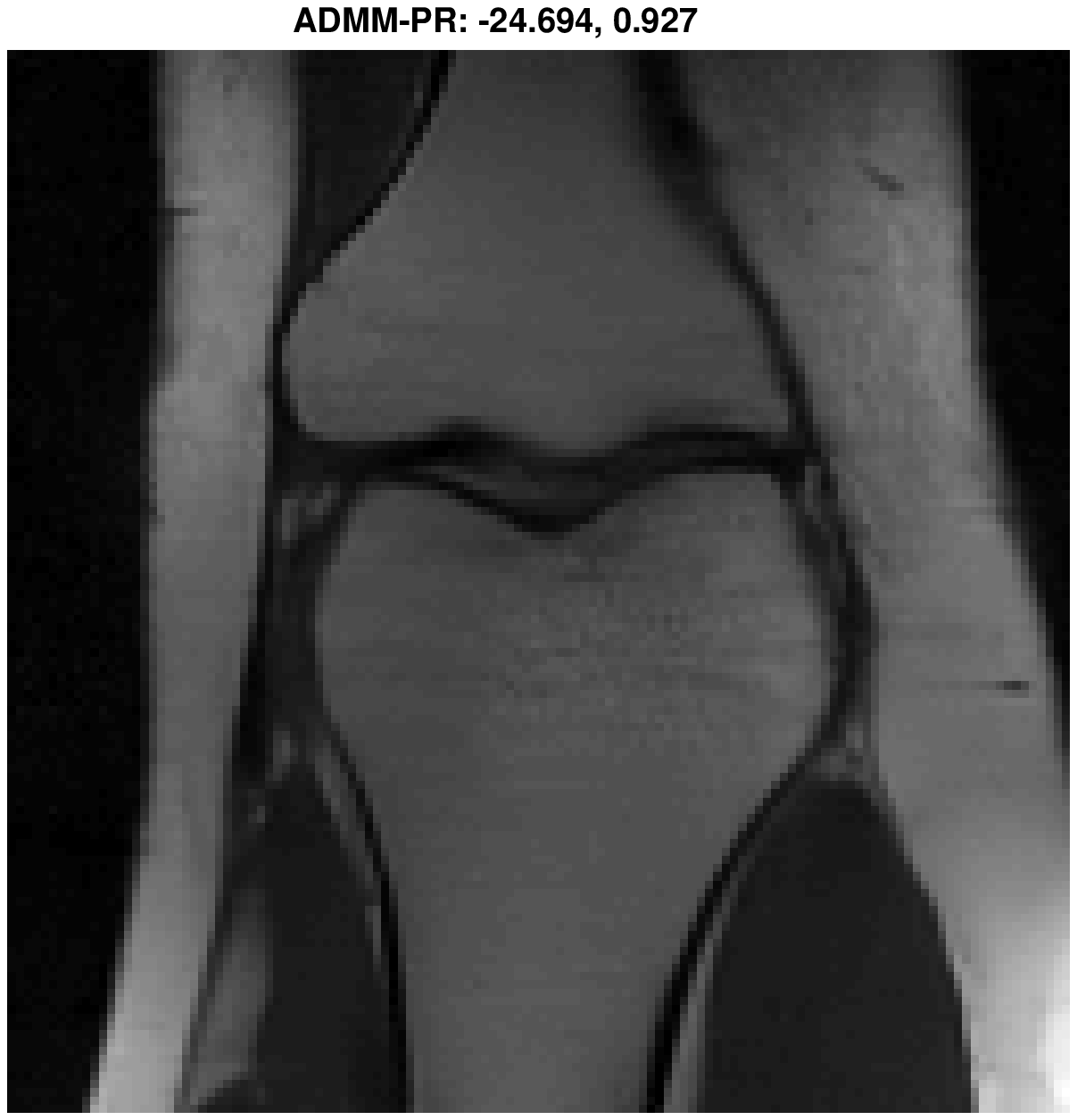}
    \end{subfigure}

    \vspace{10pt}

    \begin{subfigure}[t]{\scb\columnwidth}
        \centering
        \psfrag{BM3D-AMP-MRI: -19.263, 0.857}[b][b][\sz]{\sf \quad~~ BM3D-AMP-MRI: -19.263, 0.857}
        \includegraphics[scale=\sca,trim=0mm 0mm 0mm 0mm,clip]{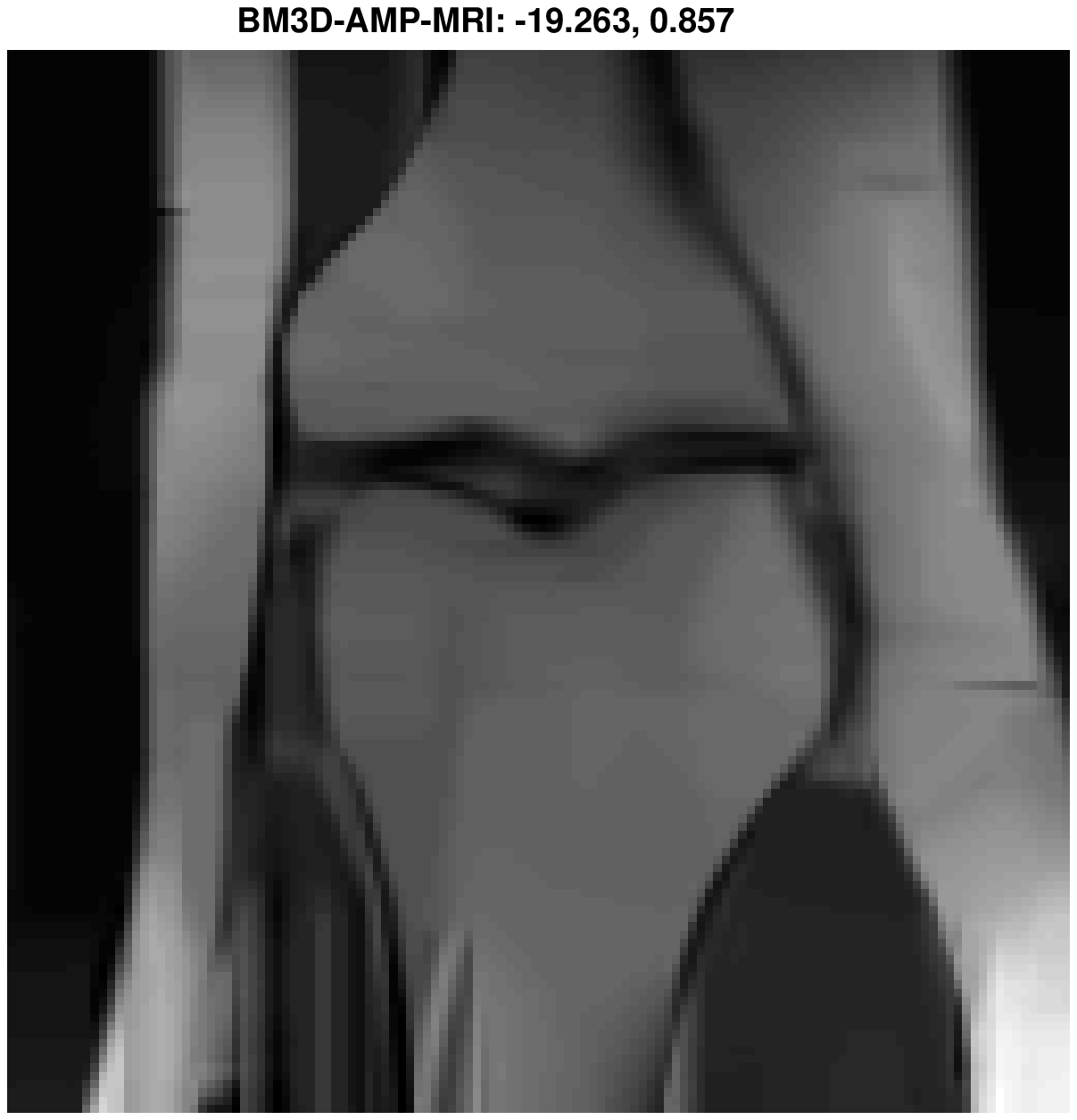}
    \end{subfigure}
    ~
    \begin{subfigure}[t]{\scb\columnwidth}
        \centering
        \psfrag{DD-VAMP: -25.097, 0.936}[b][b][\sz]{\sf \quad DD-VAMP: -25.097, 0.936}
        \includegraphics[scale=\sca,trim=0mm 0mm 0mm 0mm,clip]{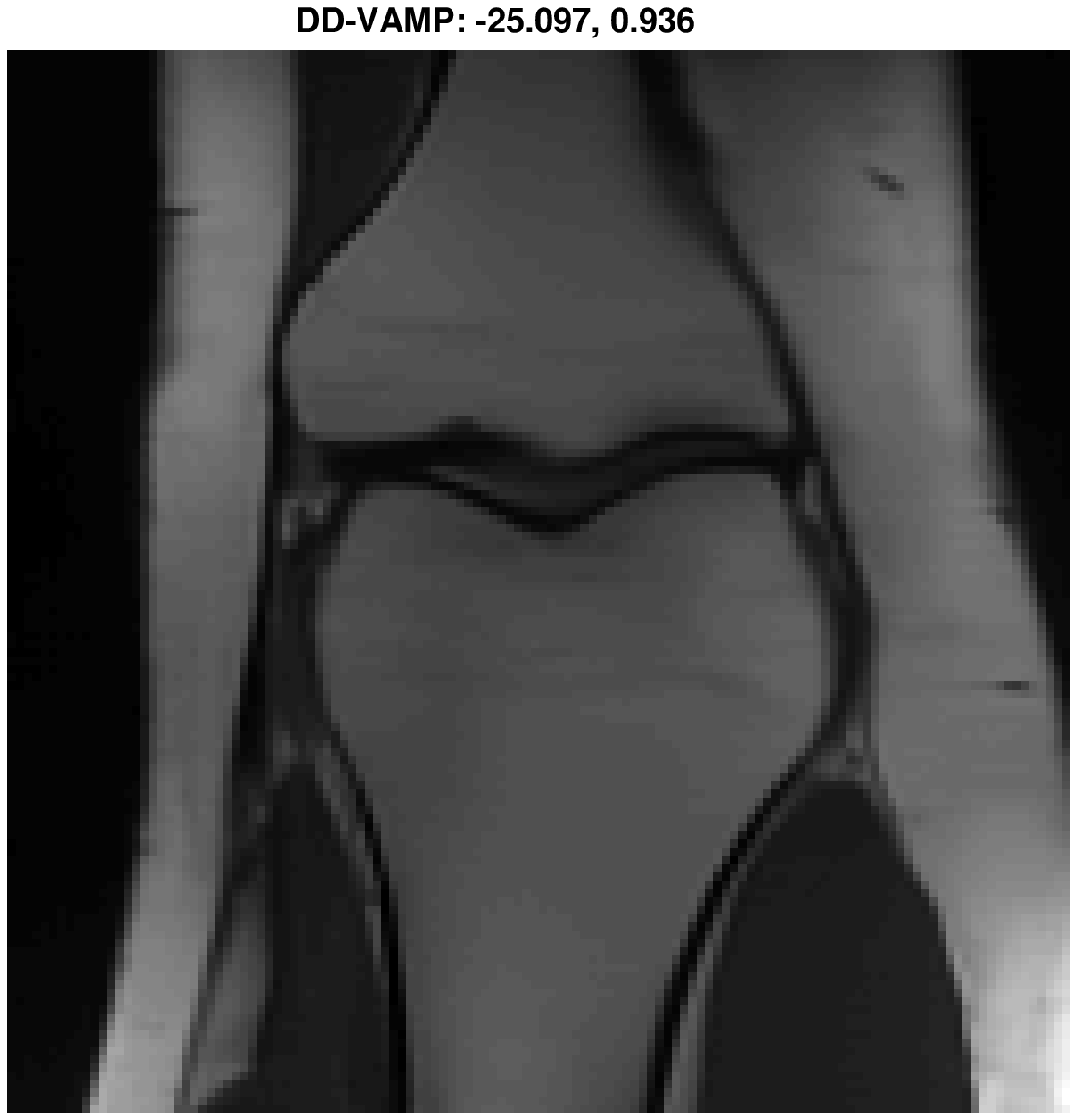}
    \end{subfigure}%
    ~
    \begin{subfigure}[t]{\scb\columnwidth}
        \centering
        \psfrag{ADMM-PR--DD-VAMP: -25.475, 0.939}[b][b][\sz]{\sf \quad~ DD-VAMP++: \textbf{-25.475, 0.939}}
        \includegraphics[scale=\sca,trim=0mm 0mm 0mm 0mm,clip]{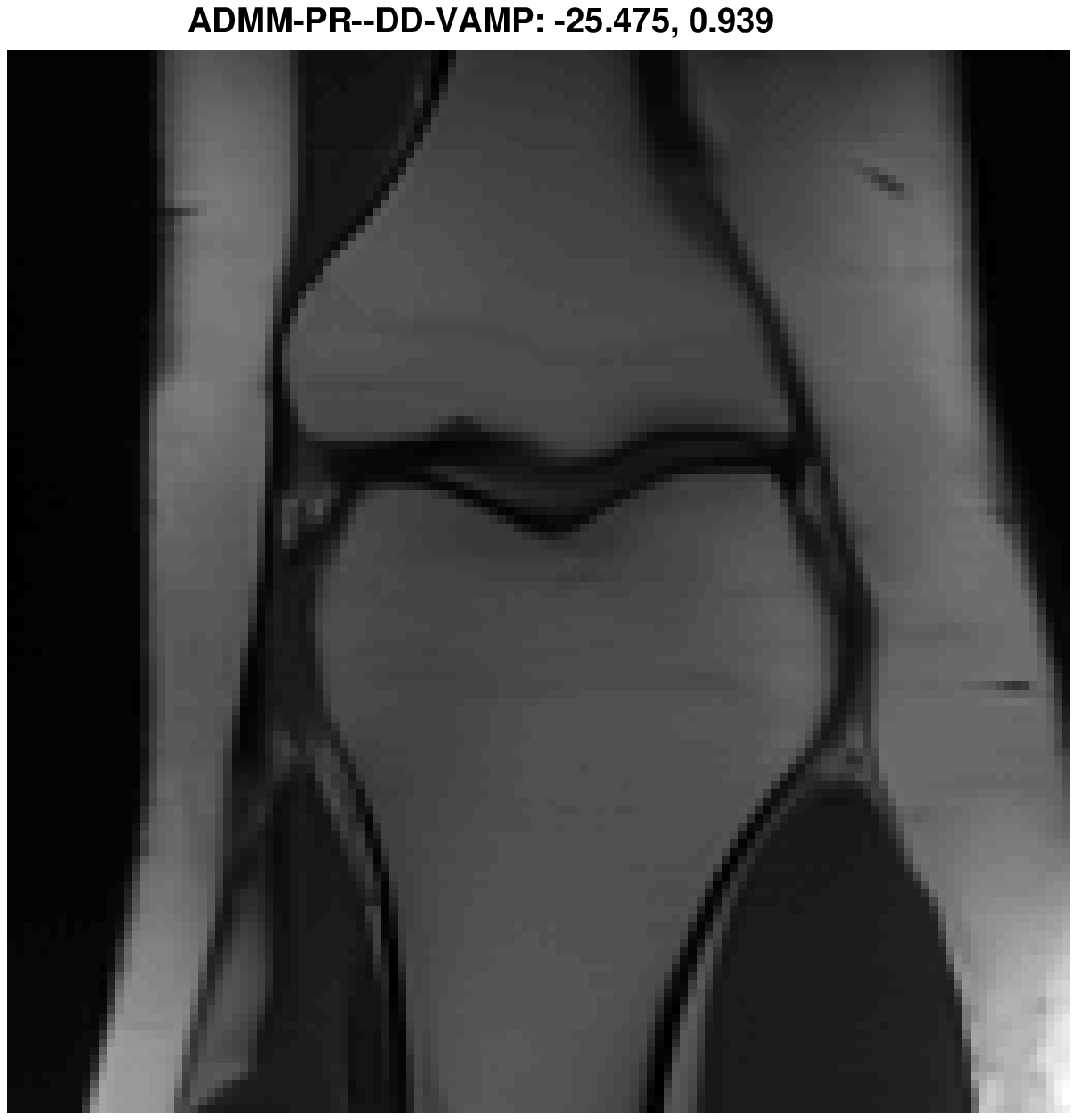}
    \end{subfigure}
    \caption{Image recovered after 150 iterations. Subplot titles identify algorithm, NMSE (dB), and SSIM.}
    \label{fig:cart_dncnn}
\end{figure}

\Figref{cart_dncnn} shows example image recoveries after 150 iterations, along with NMSE (dB) and SSIM values.
DD-VAMP++ achieved the best recovery in NMSE and SSIM. 

%

\clearpage
\bibliographystyle{IEEEtran}
\bibliography{macros_abbrev,phase,books,misc,comm,multicarrier,sparse,machine}

\end{document}